\documentstyle[epsfig]{mn}

\ifCUPmtlplainloaded \else
   \DeclareSymbolFont{UPM}{U}{eur}{m}{n}
   \SetSymbolFont{UPM}{bold}{U}{eur}{b}{n}
   \DeclareMathSymbol{\upi}{0}{UPM}{"19}
   \DeclareMathSymbol{\umu}{0}{UPM}{"16}
   \DeclareMathSymbol{\upartial}{0}{UPM}{"40}
\fi

\newcommand{\h}{$^{\rm h}$~}
\newcommand{\m}{$^{\rm m}$~}

\begin{document}

\title{A search for radio emission from Galactic supersoft X-ray sources}

\author[R.~N.~Ogley et al.]{
R.~N.~Ogley,$^{1}$\thanks{E-mail: rno@astro.keele.ac.uk}
S.~Chaty$^{2}$,
M.~Crocker,$^{3}$,
S.~P.~S.~Eyres$^{4}$,
M.~A.~Kenworthy$^{5}$,
\newauthor
A.~M.~S.~Richards$^{3}$,
L.~F.~Rodr\'{\i}guez$^{6}$ and
A.~M.~Stirling$^{4}$
\\
$^{1}$ Department of Physics, Keele University, Staffordshire ST5 5BG\\
$^{2}$ Department of Physics and Astronomy, The Open University, Walton
Hall, Milton Keynes, Buckinghamshire MK7 6AA\\
$^{3}$ University of Manchester, Jodrell Bank Observatory, Macclesfield,
Cheshire SK11 9DL\\
$^{4}$ CFA, University of Central Lancashire, Preston PR1 2HE\\
$^{5}$ Steward Observatory, 933 North Cherry Avenue, Tucson, AZ 85721, USA\\
$^{6}$ Instituto de Astronom\'{\i}a, UNAM, Campus Morelia,
Apdo. Postal 3-72, Morelia, Michoac\'an 58089, M\'exico
}

\maketitle

\begin{abstract}
We have made a deep search for radio emission from all the northern
hemisphere supersoft X-ray sources using the VLA and MERLIN
telescopes, at 5 and 8.4 GHz.  Three previously undetected sources: T
Pyx, V1974 Cygni and RX J0019.8+2156 were imaged in quiescence using
the VLA in order to search for any persistent emission.  No radio
emission was detected in any of the VLA fields down to a typical
1$\sigma$ RMS noise of 20 $\umu$Jy beam$^{-1}$, however, 17 new point
sources were detected in the fields with 5 GHz fluxes between 100 and
1500 $\umu$Jy giving an average 100 $\umu$Jy-source density of
$\sim$200 per square degree, comparable to what was found in the
MERLIN HDF survey.  The persistent source AG Draconis was observed by
MERLIN to provide a confirmation of previous VLA observations and to
investigate the source at a higher resolution.  The core is resolved
at the milliarcsec scale into two components which have a combined
flux of $\sim$ 1 mJy.  It is possible that we are detecting nebulosity
which is becoming resolved out by the higher MERLIN resolution. We
have investigated possible causes of radio emission from a wind
environment, both directly from the secondary star, and also as a
consequence of the high X-ray luminosity from the white dwarf.  There
is an order of magnitude discrepancy between observed and modelled
values which can be explained by the uncertainty in fundamental
quantities within these systems.
\end{abstract}


\section{Introduction}

Supersoft sources are a separate class of X-ray objects.  The most
popular explanation of these sources is a white dwarf \& sub-giant
companion with a high accretion rate, 100--1\,000 times greater than
in cataclysmic variables (van den Heuvel et al.\ 1992).  The large
accretion rate creates steady hydrogen burning on the white dwarf
surface causing X-ray emission.

Supersoft sources are difficult to detect in the Galaxy as the high
column density in the Galactic plane absorbs most of the soft X-ray
radiation.  Consequently there have been many more detections at high
galactic latitudes such as in the LMC, SMC and M31.  As of 1999
(Greiner 2000), there were 57 supersoft sources: 10 in the Galaxy, 4
in the SMC, 8 in the LMC and 34 in M31 and 1 in NGC 55.  Consequently,
even though sources in the LMC and SMC are further away, they have
been studied in greater detail due to their larger numbers.  For
example, in 1997 Fender, Southwell \& Tzioumis (1998) searched for
radio emission from non-Galactic supersoft sources despite them being
prohibitively further away than their Galactic counterparts.

Of the persistent super-soft sources, three have been reported to have
outflows, detected by emission lines in their optical and infrared
spectra.  What is the makeup of these sources, and are they the link
between the low velocity jets seen in star-forming systems, and the
superluminal jets seen in the microquasars?

\subsection{Sources with outflow}

Three supersoft sources have been observed with weak Doppler-shifted
optical emission lines.  The first object to be detected was
RX~J0513.9$-$6951, an LMC object which was discovered during an
outburst in 1990 (Schaeidt et al.\ 1993), and optically identified by
Pakull et al.\ (1993) and Cowley et al.\ (1993).  This source has
persistent optical emission with dips in the lightcurve every 100--200
days (Southwell et al.\ 1996).  The evidence for jets in this source
comes from strong He\,{\sc ii}, H${\beta}$ and H${\alpha}$ lines with
Doppler-shifted components at $\pm$ 4\,000 km~s$^{-1}$ (Southwell et
al.\ 1996).  These lines are still observed during an optical dip, but
with a reduced equivalent width, implying a reduction in the accretion
rate.

The second supersoft jet source to be detected was RX~J0019.8+2156
(Beuermann et al.\ 1995), a Galactic object at 2 kpc consisting of a 1
M$_{\sun}$ WD and a 1.5 M$_{\sun}$ donor star (Becker et al.\ 1998).
The optical spectrum shows strong He\,{\sc ii} emission and P~Cygni
absorption in the Balmer lines.  This source displays slower red and
blue-shifted lines with velocities of $\pm$ 815 km~s$^{-1}$, but these
lines have a high FWHM of 400 km~s$^{-1}$ (Tomov et al.\ 1998).
Quaintrell \& Fender (1998) have also found these features in infrared
spectra.  Becker et al.\ (1998) conclude that the jets in this system
come from an inclined cone of material blown off from an accretion
disc.

The third, and last, supersoft source to be discovered with an outflow
is the Galactic source RX~J0925.7$-$4758.  This was observed by Motch
(1998) over two nights in 1997.  Spectra showed prominent H${\alpha}$
blue and red-shifted lines with a velocity of 5\,800 km~s$^{-1}$.
Motch (1998) modelled the emission from a cone with a large opening
angle, at low inclination, similar to RX~J0019+2156.

The only radio detection of a supersoft source was that of the
symbiotic star AG~Draconis (Torbett \& Campbell 1987; Seaquist et al.\
1993).  Observations with the VLA showed two unresolved components
separated by $\simeq$ 1\arcsec, at fluxes of 200 and 400 $\umu$Jy,
shown in figure \ref{vla_merlin}(a).  These authors conclude that the
emission is from optically thick thermal free-free emission, based on
the majority of symbiotic systems.  However, we cannot rule out
optically thick emission in this paper.

\section{Observations}

Of the ten Galactic supersoft sources, four were observed by the VLA.
These are RX~J0019.8+2156, T~Pyx, AG~Draconis and V1974 Cygni.  Three
of these sources were observed with the VLA with the exception of AG
Dra which was observed using MERLIN for the purpose of resolving any
detail.

As we were primarily concerned with initial detections of point
sources, a standard continuum setup was used for both the VLA and
MERLIN observations.  For the VLA we used a 100 MHz bandwidth split
into two channels, and for MERLIN we used a 15 MHz bandwidth split
into 15 channels.  Individual channels were retained in order to
reduce the amount of bandwidth smearing of sources offset from the
phase-centre.

\subsection{T Pyxis}

T~Pyx is a recurrent nova first discovered by Leavitt (1914) at a
distance of 1.5 kpc (Shara 1997; Margon \& Deutsch 1998).  An
arcsecond size optical nebula around the source, which shows a large
amount of clumpy material on sub-arcsec scales , was detected from
this object (Williams 1982; Duerbeck \& Sitter 1987).  This clumpy
material would provide a rich medium for interaction with an outflow,
such as the 1\,100--1\,400 km~s$^{-1}$ jets reported by Shahbaz et
al.\ (1997) (however, this outflow probably originates in the
surrounding nebular, and not in jets, see O'Brien \& Cohen (1998) and
Margon \& Deutsch (1998)).

Observations were taken using the VLA in CnB configuration on 2000
March 3, 0415-0615 UT at 8.4 GHz and in D configuration on 2000
September 26 \& 28, 1600-1710 \& 1400-1630 UT respectively.  Data were
calibrated using the standard flux calibrator 3C~147, and phase
calibrator PKS~J0921$-$2618.  No emission was observed from either the
source, or any of the 7 surrounding sources from the SIMBAD database.
We can put a $1 \sigma$ RMS upper-limit of 22 $\umu$Jy beam$^{-1}$ at
4.8 GHz, and 19 $\umu$Jy beam$^{-1}$ at 8.4 GHz.

An observing log, together with beam sizes, RMS noise and source
details for all sources is given in table \ref{observing log}.

\begin{table}
\caption{Observation log and source details.  Please note that the
Merlin observations are only at 5 GHz and the two fluxes presented are
for two components in the image.}
\label{observing log}
\begin{tabular}{lll}
\hline
		& \multicolumn{2}{c}{T Pyx}\\
\hline
Band		& 4.8 GHz (C) 		& 8.4 GHz (X) 		\\
Telescope 	& VLA D 		& VLA CnB 		\\
Date (2000)	& Sep 26 1600--1710 	& Mar 03 0415--0615 	\\
		& Sep 28 1400--1630 	&  			\\
MJD		& 51813			& 51606			\\
		& 51815			&			\\
Flux cal 	& \multicolumn{2}{c}{3C 147} 			\\
Phase cal 	& \multicolumn{2}{c}{PKS J0921$-$2618} 		\\
$1 \sigma$ RMS 	& 22 $\umu$Jy/beam 	& 19 $\umu$Jy/beam	\\
Beam size 	& 28\farcs9 $\times$ 10\farcs40 		&
	    	   2\farcs12 $\times$  2\farcs04		\\
Beam angle	& $-3\fdg98$ 		& 47\fdg6		\\

\hline
		& \multicolumn{2}{c}{V1974 Cygni} 		\\
		& \multicolumn{2}{c}{($\equiv$ Nova Cygni 1992)}\\
\hline
Band		& 4.8 GHz (C)		& 8.4 GHz (X)		\\
Telescope	& VLA DnC		& VLA CnB		\\
Date (2000)	& Jun 27 0800--0945 	& Mar 09 1500--1600 	\\
		&			& Mar 10 1330--1430 	\\
MJD		& 51722			& 51612			\\
		&			& 51613			\\
Flux cal	& \multicolumn{2}{c}{3C 286} 			\\
Phase cal	& \multicolumn{2}{c}{Ohio W 538} 		\\
$1 \sigma$ RMS	& 24.6 $\umu$Jy/beam	& 19.8 $\umu$Jy/beam 	\\
Beam size	& 13\farcs8  $\times$ 8\farcs03  		&
		   2\farcs12 $\times$ 0\farcs92 		\\
Beam angle	& 88\fdg9		& $-79\fdg 83$ 		\\

\hline
		& \multicolumn{2}{c}{RX J0019.8+2156} 		\\
		& \multicolumn{2}{c}{($\equiv$ QR And)} 	\\
\hline
Band		& 4.8 GHz (C)		& 8.4 GHz (X) 		\\
Telescope	& VLA DnC		& VLA CnB		\\
Date (2000)	& Jun 27 1300--1530 	& Mar 11 1810--2050 	\\
MJD		& 51722			& 51614			\\
Flux cal	& \multicolumn{2}{c}{3C 48} 			\\
Phase cal	& \multicolumn{2}{c}{PKS J0010+1724} 		\\
$1 \sigma$ RMS	& 22.7 $\umu$Jy/beam	& 19.8 $\umu$Jy/beam 	\\
Beam size	& 13\farcs3  $\times$ 8\farcs12			&
		   2\farcs19 $\times$ 0\farcs95			\\
Beam angle	& 76\fdg 6		& $-88\fdg 68$		\\

\hline
		& \multicolumn{2}{c}{AG Dra} \\
\hline
Band		& 5.0 GHz (C)		&			\\
Telescope	& MERLIN		&			\\
Date (2000)	& Mar 18 1010--1130	& 			\\
		& Mar 25 0630--2140	&			\\
		& Mar 26 0510--2250	&			\\
MJD		& 51621			& 			\\
		& 51628			&			\\
		& 51629			&			\\
Flux cal	& \multicolumn{2}{c}{3C 286 \& OQ 208}		\\
Phase cal	& \multicolumn{2}{c}{7C 1603+6954}		\\
Peak Fluxes	& 570 $\umu$Jy		& 420 $\umu$Jy	 	\\
$\alpha_{\rm J2000}$	& 16$^{\rm h}$ 01$^{\rm m}$ 41$\fs$007	& 
			  16$^{\rm h}$ 01$^{\rm m}$ 41$\fs$046	\\
$\delta_{\rm J2000}$	& 66\degr ~48\arcmin ~~10\farcs13	&
			  66\degr ~48\arcmin ~~10\farcs46	\\
$1 \sigma$ RMS	& 103 $\umu$Jy/beam	&			\\
Beam size	& \multicolumn{2}{l}{48.9 $\times$ 68.5 mas} 	\\
Beam angle	& $-27\fdg9$		&			\\
\hline

\end{tabular}
\end{table}

\subsection{V1974 Cygni ($\equiv$ Nova Cygni 1992)}

V1974 Cygni was extensively studied at all wavelengths following an
outburst in 1992.  Complementary VLA and MERLIN images show that the
source initially flared then decayed, as a shell from the outburst
expanded (Hjellming 1992; Eyres et al.\ 1996).  They concluded that
the emission from the nova was thermal and initially optically thick,
then gradually became optically thin.  This is expected from an
expanding shell of material.

This source was observed by the VLA in CnB configuration on 2000 March
09 \& 10 at 1500-1600 and 1330-1430 UT respectively at 8.4 GHz and in
DnC configuration on 2000 June 27 at 0800-0945 UT at 4.8 GHz. Data
were calibrated using the standard flux calibrator 3C~286 and
phase-referenced using Ohio~W~538.  No emission was detected from Nova
Cygni 1992.  We can put a $1 \sigma$ RMS upper limit to any emission
of 24.6~$\umu$Jy~beam$^{-1}$ at 4.8~GHz, and 19.8~$\umu$Jy~beam$^{-1}$
at 8.4~GHz.

\subsection{RX J0019.8+2156 ($\equiv$ QR Andromedae)}

RX J0019.8+2156 was observed by the VLA in CnB configuration on 2000
March 11 at 1810-2050 UT at 8.4 GHz and in DnC configuration on 2000
June 27 at 1300-1530 UT at 4.8 GHz.  The flux calibrator used was
3C~48 and the data were phase referenced using PKS~J0010+1724.  No
emission was detected from RX J0019.8+2156.  We can therefore put a $1
\sigma$ RMS upper limit to any emission of 22.7~$\umu$Jy~beam$^{-1}$
at 4.8~GHz, and 19.8~$\umu$Jy~beam$^{-1}$ at 8.4~GHz.

\subsection{AG Draconis}

\begin{figure*}
\epsfig{file=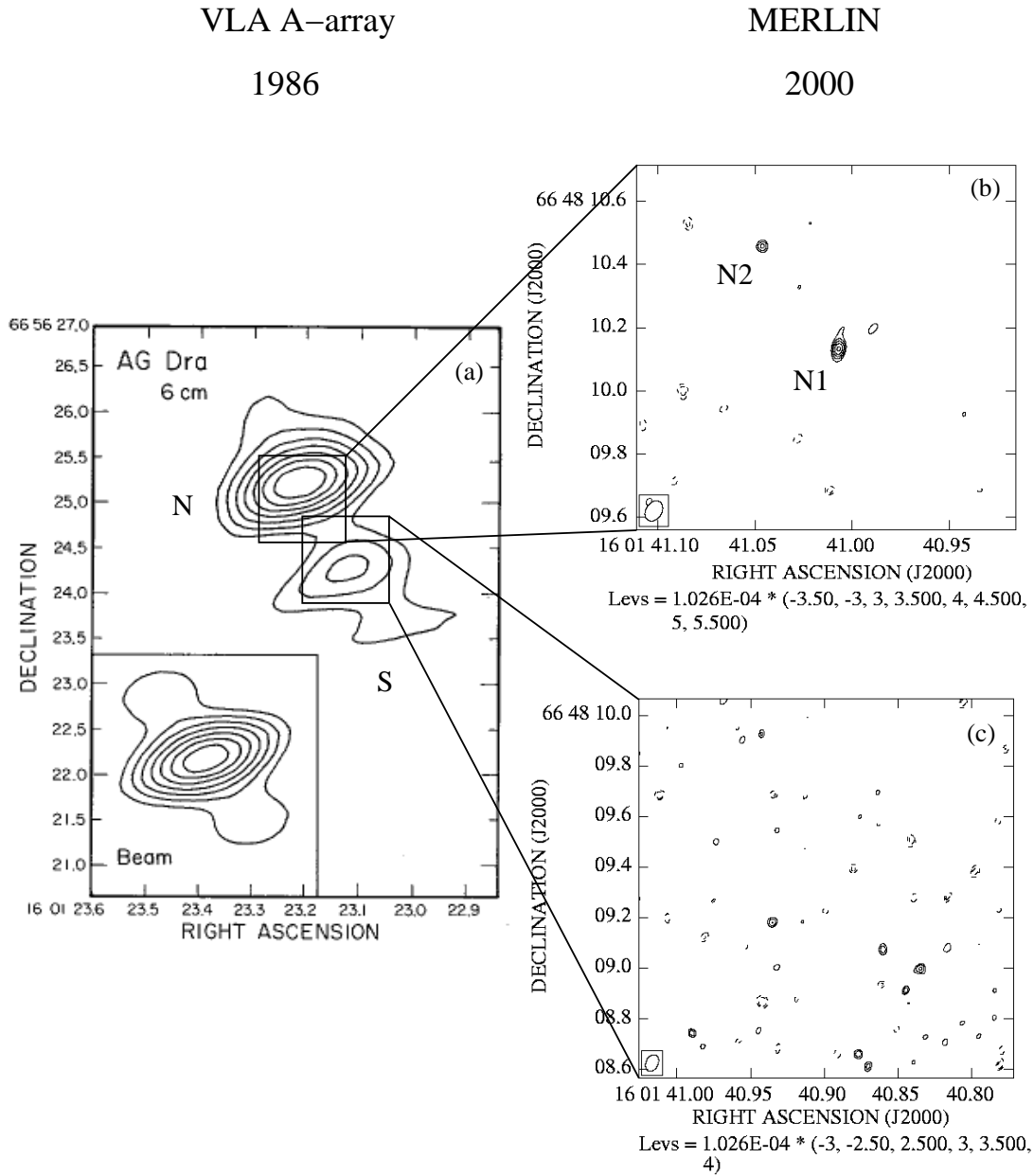, width=6in}
\caption{(a) Previous VLA observations of AG Dra. (Torbett \& Campbell
1987) (b) Higher resolution MERLIN observations at 4.994 GHz.  The
core is clearly detected and resolved into two components at the 48.9
$\times$ 68.5 mas resolution (as shown by the beam in the lower left
corner).  Any extension in component N1 is probably due to noise in
the image.  Fluxes of components N1 and N2 are 570 and 420 $\umu$Jy
respectively.  It is possible that we are resolving out any extended
nebulosity around the core in the MERLIN image.  (c) MERLIN
observations of the southern VLA component with lower contours to
bring out the noise in the image.  There are no clear detections of
emission.}
\label{vla_merlin}
\end{figure*}

To confirm the previous VLA detection, shown in figure
\ref{vla_merlin}(a), we observed AG Dra on 2000 March 18, 25 \& 26 at
1010--1130, 0630--2140 and 0510--2250 UT respectively with MERLIN.
Due to the superior resolving power of MERLIN over the VLA we intended
to resolve the two unresolved components previously detected, and
obtain some information about the source's structure.  The MERLIN
image has a beam size of 48.9 $\times$ 68.5 mas at an angle of $-27.9
\degr$, and a $1 \sigma$ RMS noise of 103~$\umu$Jy~beam$^{-1}$ at
4.994~GHz.

The core component is clearly detected in the MERLIN image and is
resolved into two point sources at $\alpha_{\rm J2000}$ = 16\h 01\m
41\fs007, $\delta_{\rm J2000}$ = +66\degr ~48\arcmin ~10\farcs13 with
a flux of 570 $\umu$Jy (designated as N1 in figure
\ref{vla_merlin}(b)), and $\alpha_{\rm J2000}$ = 16\h 01\m 41\fs046,
$\delta_{\rm J2000}$ = +66\degr ~48\arcmin ~10\farcs46 and a flux of
420 $\umu$Jy (designated as N2).  These fluxes imply a lower limit of
$\sim$8000 K for the brightness temperature, consistent with the
interpretation of the emission being optically thick free-free
emission, however, we do not rule out optically thin emission in this
paper.  We note that the extension to the north of the component N1 is
probably due to noise in the image.

The southwest component in the VLA image is not so clearly defined.
If the source has not increased in flux between the epochs, and is a
point source then it will be indistinguishable from the noise in the
MERLIN image (shown in figure \ref{vla_merlin}(c)).  We do observe a
number of possible point sources with fluxes around 300--350~$\umu$Jy,
but these are ambiguous compared to the noise.  We have lowered the
contours compared to figure \ref{vla_merlin}(b) to show the noise in
more detail.

\subsection{New objects}

All three VLA fields were imaged with a diameter of 10.2 arcmin in
order to check for emission from known sources, and to detect new
objects in the fields.  However, interferometric images all suffer
from bandwidth smearing or chromatic aberration.  This is particularly
important when the observed bandwidth is large and a wide field is
imaged.  The effects of these are to smear sources in a radial
direction from the phase centre, and to reduce the observed peak flux.
The consequence of this effect is that the signal to noise ratio
increases with off-axis beam angle.

For an observation taken with a bandwidth of 50 MHz at 8.3 GHz, and
with a circular beam 2 arcsec ~in diameter, a point source will suffer
a 10 per cent reduction in flux at a distance of 3 arcmin 50 arcsec
from the phase centre, and a 50 per cent reduction at a distance of
around 11 arcmin, this situation becomes more severe with a smaller
beam  (see Bridle \& Schwab (1989) for details).

We have uploaded to the CDS service for Astronomical
Catalogues\footnote{http://cdsweb.u-strasbg.fr/cats/Cats.htx} a table
which lists the sources taken from the Simbad catalogue together with
our limits to the radio emission from the source based on the
bandwidth smeared $5 \sigma$ RMS noise.  While none of the known
sources were detected, we observed 17 new sources with a flux greater
than $5 \sigma$.  An uploaded table as above lists these new sources,
together with a Gaussian-fitted flux based on the image beam, and true
flux due to the bandwidth smearing in the image, (at the bands $C$ and
$X$ given in table \ref{observing log}).  A spectral index is also
calculated from the bandwidth smeared fluxes and follows the
convention of $S \propto \nu^{\alpha}$.

We can therefore estimate that the density of sources with flux $>$
100 $\umu$Jy is about 200 sources per square degree at 5 GHz, broadly
consistent with limits from the deep radio survey by MERLIN of the
Hubble Deep Field (Muxlow et al. 1999).  The spectral indices of these
sources are generally negative or consistent with a negative value,
suggesting that most are non-thermal extragalactic objects, however
caution has to been drawn to this conclusion since data at the two
frequencies is not simultaneous, and there is a difference in angular
resolution between the two frequencies.

\section{Wind emission}

In this section we describe simple analytical models for wind emission
and apply them to our sample.  Initially we are not interested in
where the wind comes from or what it consists of, rather that a
generic wind with a particular composition, being ejected at a
particular rate from a source at a given distance from us, would
produce radio emission.  The optically thin flux from a source with
these parameters can be calculated and is given by
\begin{equation}
S = 23.2 \left( \frac {\dot{M}}{\mu v_{\rm km/s}}\right)^{4/3} \left(\nu
\gamma g Z^2 \right)^{2/3} D^{-2}\;\rm{Jy}
\label{wind_fluxes}
\end{equation}
where $\dot{M}$ is in M$_\odot$ yr$^{-1}$, $\mu$ is the mean atomic
weight, $v_{\rm km/s}$ is the wind velocity in km s$^{-1}$, $\nu$ is
in Hz, $\gamma$ is the ratio of electron to ion number density, $g$ is
the Gaunt factor, $Z$ is the ionic charge and $D$ is the distance in
kpc (Wright \& Barlow 1975).  The Gaunt factor here has a slight
dependence on temperature and composition and if one does not consider
extremely hot winds, then it can be approximated by
\begin{equation}
g = 9.77 + 1.27 \log_{10}{\left(T^{3/2}/\nu Z\right)}
\end{equation}
(Leitherer \& Robert 1991).

We can now investigate in more detail from where the wind originates,
what is its expected composition and what is the expected radio
emission from such a region.

\subsection{Generic secondary wind}

To create sustained mass transfer between the binary components either
the secondary\footnote{In this paper we use the convention that the WD
is designated the primary object and the companion, or donor, is
designated the secondary.  Subscripts 1 and 2 are used in mass and
radius relationships to distinguish between these two objects} is
undergoing Roche lobe overflow, the wind is providing the accretion
via Bondi-Hoyle accretion, or both.  Which scenario occurs depends on
the nature of the secondary, and the binary parameters.

The symbiotic system AG Dra with a $\sim 1.5$ M$_{\sun}$ K-giant, a
$\sim 0.5$ M$_{\sun}$ white dwarf and a period of 544 days does not
normally fill its Roche lobe (the stellar radius of the K-giant is
$\sim 30$ R$_{\sun}$ and the Roche lobe size is $\sim 170$
R$_{\sun}$).  Since the system must be accreting then the logical
conclusion is that the secondary is on the Asymptotic Giant Branch
(AGB) and feeding the primary by a slow, dense wind.  Out of the four
sources considered here, this is the most likely to be feeding its
environment directly by a wind.  Calculating the expected flux from
this wind uses a temperature, velocity, outflow and distance of $T$ =
15\,000 K, $v$ = 30 km s$^{-1}$, $\dot{M}$ = 1.2--1.7 $\times 10^{-8}$
M$_{\sun}$ yr$^{-1}$ and $D$ = 1.56--1.81 kpc from Tomova \& Tomov
(1999), compositions of $\mu = 1.4$, $\gamma = 1$ and $Z = 1$ from
Nussbaumer \& Vogel (1987) and Spergel, Giuliani \& Knapp (1983).
These parameters yield a flux at 5 GHz of $S_{C}$ = 12--26 $\umu$Jy.

In contrast to AG Dra, the three other sources either have a small
secondary companion and therefore Roche lobe overflow would be
providing the mass-transfer, or as in the case for V1974 Cyg, radio
emission would have faded and dissipated from the time of the last
outburst in 1992.

\subsection{Evaporated secondary wind}

In the previous section it was shown that only the AGB binary AG Dra
would have a strong enough secondary wind to emit significant radio
emission.  However, if the WD is radiating sufficient soft X-rays then
a corona will be formed around the secondary star, which causes a
strong stellar wind to be evaporated (Basko \& Sunyaev 1973).  This
type of interaction has been investigated in the context of the
supersoft sources by van Teeseling \& King (1998, hereafter vTK98) and
they show that the mass loss rate in the secondary due to irradiation
is
\begin{equation}
\dot{M}_{7} \simeq -3  \phi \frac{r_2}{a_{9}} \left(
m_2 L_{30} \eta_{s} \right)^{1/2} \; {\rm M_{\odot}\;yr^{-1}}
\label{mass-loss-equation}
\end{equation}
where $\phi$ is an irradiation efficiency factor, $r_2$ is the radius
of the secondary in solar units, $a_{9}$ is orbital separation in
units of $10^{9}$ m, $m_2$ is the mass of the secondary in solar
units, $L_{30}$ is the irradiation luminosity in units of $10^{30}$ J
s$^{-1}$ ($\equiv 10^{37}$ erg s$^{-1}$), $\eta_s$ is an ionisation
efficiency parameter, and $\dot{M}_{7}$ is given in units of
$10^{-7}$.  It is reasonable, therefore, to assume that since the
supersoft sources produce soft X-rays during quiescence from their
sustained nuclear burning, that the majority (if not all) of the
supersoft sources have mass losses which are wind dominated.

To obtain an estimate for the mass loss from the secondary due to
irradiation from the WD, some assumptions to the geometry of the
system need to be made.  In equation \ref{mass-loss-equation}, the
variable $\phi$ is an efficiency parameter representing the fraction
of the companion's face which is being radiated and the fraction of
the wind mass escaping the system.  Commentary on the importance of
$\phi$ is given in Knigge, King \& Patterson (2000), and a value of 1
is expected.  The other unknown parameter in equation
\ref{mass-loss-equation} is the efficiency of the incident radiation
in driving a wind, $\eta_{s}$.  For a wide range of temperatures,
including the expected temperature of the irradiation (few $\times
10^{5}$ K), $\eta_{s} \simeq 1$.  We can therefore calculate an
estimation for the mass-loss in the wind from the secondary, which is
given in table \ref{system_parameters}.
\begin{table*}
\caption{System parameters and radio fluxes for the supersoft sources.}
\label{system_parameters}
\begin{minipage}{6in}
\begin{tabular}{llllll}
\hline
		& \multicolumn{2}{c}{RX J0019} & AG Dra & V1974 Cyg & T Pyx \\
\hline
Secondary 	& F V
		& M V     
		& K III\footnote{Tomov, Tomova \& Ivanova (2000)} 
		& M5 V
		& M V\footnote{Szkody \& Feinswog (1988)}
\\
Period		& 15.8 h\footnote{Will \& Barwig (1996)}
		&
		& 554 d\footnote{Miko{\l}ajewska et al.\ (1995)}
		& 1.95 h\footnote{Skillman et al.\ (1997)}
		& 1.8 h\footnote{Schaefer et al.\ (1992)}
\\
$m_1$		& 0.75\footnote{Meyer-Hofmeister, Schandl \& Meyer (1997)}
		&
		& 0.4--0.6$^{d}$
		& $0.9 \pm 0.2$\footnote{Retter, Leibowitz \& Ofek (1997)}
		& 1.2\footnote{Contini \& Prialnik (1997)}
\\
$m_2$		& 1.5$^{g}$
		& 0.31--0.44\footnote{Deufel et al.\ (1999)}	   
		& $\sim 1.5^{d}$
		& $0.23 \pm 0.08$ $^{i}$
		& 0.12\footnote{Clemens et al.\ (1998);
				Patterson (1998)}
\\
$r_2$		& 1.3
		& 0.36--0.49
		& 28--32$^{d}$
		& $0.27 \pm 0.08$
		& 0.17\footnote{Paczy\'{n}ski (1971)}
\\
$L_{30}$	& 0.3--0.9\footnote{Beuermann et al.\ (1995)}
		&
		& 0.14\footnote{Greiner et al.\ (2000)}
		& 2\footnote{Kahabka \& van den Heuvel (1997)}
		& 0.2\footnote{Patterson et al.\ (1998)}
\\
$a_9$		& 2.9
		& 1.79--1.87	& 249	& $0.57 \pm 0.04$ & 0.57	\\
$\dot{M}_{7}$	& 1.3 & 0.18--0.52	& 0.17	& $1.0 \pm 0.5$	& 0.14	\\
$D$ (kpc)	& 2.1$^{m}$
		&
		& 1.7$\pm$0.1$^{a}$
		& 1.8\footnote{Chochol et al.\ (1993); Rosino et al.\ (1996)}
		& $3.5 \pm 1.0^{p}$
\\
$T$		& $5 \times 10^{5}$\footnote{van Teeseling \& King (1998)}
		&
		& $5 \times 10^{5}$ $^{r}$
		& $5 \times 10^{5}$ $^{r}$
		& $5 \times 10^{5}$ $^{r}$
\\
$v_{\rm km/s}$	& 60
		&
		& 60
		& 60
		& 60
\\
$R_{\rm C}$ (R$_{\sun}$)	& 510 & 140--280& 130	& $430 \pm 150$	& 120	\\
$R_{\rm X}$ (R$_{\sun}$)	& 360 & 95--190 & 90	& $300 \pm 110$	& 80	\\
$S_{\rm C}$(theoretical secondary) ($\umu$Jy)
				& & & 12--26 & & \\
$S_{\rm C}$(theoretical evaporated) ($\umu$Jy)		& 110 & 7.5--31	& 11	& $110 \pm 70$	& $2.6 \pm 1.4$	\\
$S_{\rm X}$(theoretical evaporated) ($\umu$Jy)		& 140 & 10--43	& 15	& $160 \pm 100$	& $3.5 \pm 1.9$	\\
$S_{\rm C}$(observed) ($\umu$Jy) & $< 100$ & & $\sim$ 1000 & $< 120$ & $< 110$ \\
$S_{\rm X}$(observed) ($\umu$Jy) & $< 80$  & & --  & $< 100$ & $< 95$  \\
\hline
\end{tabular}
\end{minipage}
\end{table*}

Together with the mass-loss rate and equation \ref{wind_fluxes}, the
flux from the wind can be found.  Unfortunately, the expected fluxes are
not as precise as one hopes as a large number of important parameters
such as composition ($\mu$, $\gamma$ and $Z$), wind temperature and
velocity ($T$ and $v_{\infty}$) are unknown.

van Teeseling \& King (1998) consider He\,{\sc ii} to be the main
absorber of the WD radiation, which has implications to the
composition and temperature of the resulting wind.  The temperature of
the wind will range from the ionisation temperature of He\,{\sc ii} to
the temperature of the incident radiation, so we use a wind
temperature of $T \simeq 5 \times 10^{5}$ in this paper.  The
composition of a purely He\,{\sc ii} wind would give values for the
mean atomic weight $(\mu = 4)$, the electron to ion number density
ratio $(\gamma = 2)$ and the ionic charge $(Z = 2)$.  It would be
surprising if these were realistic values for the wind, but they are
useful approximations and only enter the wind-flux equation
(\ref{wind_fluxes}) as $S \propto \mu^{-4/3}\gamma^{2/3}Z^{4/3}$ and
so, small changes would not have a large effect on the estimation.
The final variable to estimate is the wind velocity, $v_{\infty}$.
vTK98 approximate the velocity to be 0.3 times the isothermal sound
speed.  For sound speeds of the order $\sim 100$ km s$^{-1}$ this
would give approximate wind velocities of $\sim 30$ km s$^{-1}$.
Using these approximations to the wind gives flux estimates at 5 GHz
($S_{\rm {C}}$) and 8.3 GHz ($S_{\rm{X}}$) shown in Table
\ref{system_parameters}.

\subsection{White dwarf wind}

We have tried to estimate the radio flux from a white dwarf wind, but
the uncertainties are too large to confidently quote numbers.  Both
Bondi-Hoyle and Roche lobe overflow accretion cannot be reliably used
to predict the mass-transfer rate, and therefore the amount of excess
material that the white dwarf can process is also undetermined.

\subsection{Observational comments to wind emission}

From our MERLIN and VLA observations we can provide the flux for AG
Draconis and upper limits to the fluxes of the other three supersoft
sources in this paper (presented in table \ref{system_parameters}).
Given the spread in the observable system parameters, the expected
flux for all three non-detected sources is at, or below the 5 $\sigma$
cutoff for detections.  The surprise in this analysis is with AG Dra
where we detect with MERLIN two unresolved point sources with a
combined flux of $\sim$ 1 mJy.  This is much larger than both the wind
flux and an evaporated wind at that frequency of around 10-20 $\umu$Jy
and so we can conclude either the assumptions used in these models are
unrealistic, or a different emission mechanism is producing the
observed radio flux.

It could be significant that the only detected fluxes originate from a K
giant secondary rather than F or M dwarfs.  The geometry and accretion
in AG Dra is not expected to be from Roche lobe overflow, but rather
from the wind of the AGB secondary.

\section{Conclusions}

We have searched for quiescent, persistent radio emission from the
northern hemisphere supersoft X-ray sources, and taken a high
resolution image of the one known source with emission.  

We have improved the radio positions, source size and flux from the
persistent emitter AG Dra with MERLIN.  The core is resolved at the
milliarcsec scale into two components with a combined flux of $\sim$
1000 $\umu$Jy.  It is possible that the core detected in the VLA image
has significant nebulosity and is resolved out by the higher MERLIN
resolution.  A possible south component detected by the VLA was
unconfirmed, although this could be due to it having a flux lower than
the noise in the MERLIN image, or also being resolved out.

No new emission was detected from RX J0019.8+2156, T Pyx and V1974
Cygni down to a 1$\sigma$ RMS noise level of around 20 $\umu$Jy beam$^{-1}$.
No emission was detected from any previously known Galactic or
extragalactic source in the 3, 104 square arcsec fields imaged, and we
place upper-limits to radio emission from these sources.

We have investigated possible causes of radio emission from a wind
environment, both directly from the secondary star, and also as a
consequence of the high X-ray luminosity from the WD.

A total of 17 new point-sources were imaged with fluxes between 100
and 1500 $\umu$Jy giving an estimate to the density of sources with a
flux $>$ 100 $\umu$Jy of 200 per square degree, at 5 GHz, in this
region of the sky.

\section{acknowledgments}

RNO wishes to thank the hospitality of the MERLIN national facility at
the Jodrell Bank Observatory, especially Tom Muxlow.  SC acknowledges
support from grant F/00-180/A from the Leverhulme Trust.

The National Radio Astronomy Observatory is a facility of the National
Science Foundation operated under cooperative agreement by Associated
Universities, Inc.  MERLIN is a national facility operated by the
University of Manchester on behalf of PPARC.  This research has made
use of the SIMBAD database, operated at CDS, Strasbourg, France

\end{document}